\documentclass[twocolumn,prb,aps,amssymb,superscriptaddress,showpacs]{revtex4-1}
\usepackage[T1]{fontenc}
\usepackage{amsmath}
\usepackage{amssymb}
\usepackage{graphicx}
\usepackage[colorlinks=true,citecolor=blue,linkcolor=blue,urlcolor=blue]{hyperref}%
\usepackage{epsfig}
\usepackage{dcolumn}
\usepackage{bm}
\usepackage{float}

\def\be{\begin{equation}}
\def\ee{\end{equation}}
\def\bea{\begin{eqnarray}}          
\def\eea{\end{eqnarray}}
\def\bi{\begin{itemize}}
\def\ei{\end{itemize}}

\begin{document}

\title{ 
Kibble-Zurek mechanism with a single particle:  \\
dynamics of the localization-delocalization transition in the Aubry-Andr\'e model
}

\author{Aritra Sinha}

\affiliation{Jagiellonian University, Marian Smoluchowski Institute of Physics,\\ 
             {\L}ojasiewicza 11, PL-30348 Krak\'ow, Poland}

\author{Marek M. Rams}
\affiliation{Jagiellonian University, Marian Smoluchowski Institute of Physics,\\ 
             {\L}ojasiewicza 11, PL-30348 Krak\'ow, Poland}

\author{Jacek Dziarmaga}
\affiliation{Jagiellonian University, Marian Smoluchowski Institute of Physics,\\ 
             {\L}ojasiewicza 11, PL-30348 Krak\'ow, Poland}

\date{\today}

\begin{abstract}
The Aubry-Andr\'e 1D lattice model describes a particle hopping in a pseudo-random potential. Depending on its strength $\lambda$, all eigenstates are either localized ($\lambda>1$) or delocalized ($\lambda<1$). Near the transition, the localization length diverges like $\xi\sim(\lambda-1)^{-\nu}$ with $\nu=1$. We show that when the particle is initially prepared in a localized ground state and the potential strength is slowly ramped down across the transition, then -- in analogy with the Kibble-Zurek mechanism -- it enters the delocalized phase having finite localization length $\hat\xi\sim\tau_Q^{\nu/(1+z\nu)}$. Here $\tau_Q$ is ramp/quench time and $z$ is a dynamical exponent. At $\lambda=1$ we determine $z\simeq2.37$ from the power law scaling of energy gap with lattice size $L$. Even though for infinite $L$ the model is gapless, we show that the gap relevant for excitation during the ramp remains finite. Close to the critical point it scales like $\xi^{-z}$ with the value of $z$ determined by the finite size scaling. It is the gap between the ground state and the lowest of those excited states that overlap with the ground state enough to be accessible for excitation. We propose an experiment with a non-interacting BEC to test our prediction. Our hypothesis is further supported by considering a generalized version of Aubry-Andr\'{e} model possessing an energy-dependent mobility edge.
\end{abstract}

\maketitle
\section{Introduction}

Disordered systems have generated decades of intense research following a seminal work by Anderson in 1958~\cite{Anderson1958}. It showed that the presence of sufficiently strong disorder can halt the mobility of electrons in a lattice. About two decades later, the famous \textit{Gang of Four} paper~\cite{G41979} presented the \textit{Scaling Theory of Localization}, predicting, that all eigenstates for non-interacting electrons in dimensions $d \leq 2$ are localized. On the other hand, when $d > 2$, there is a disorder-induced localization transition, indicating, that below a certain disorder strength the lattice can also support eigenstates extended through it. How this simple picture is modified in the presence of interactions, was explored relatively recently by Basko, Aleiner and Altshuler ~\cite{Basko2006}. They showed, that an interacting many-body system can undergo a so-called many-body localization (MBL) transition in the presence of quenched disorder. This initiated a new effort to study quantum statistical mechanics of localization models, specifically investigating the fate of many-body quantum systems in a wide gamut of circumstances --- as to whether it will thermalize or stay localized.

\begin{figure}[t]
\begin{center}
   \includegraphics[width=0.95\columnwidth,clip=true]{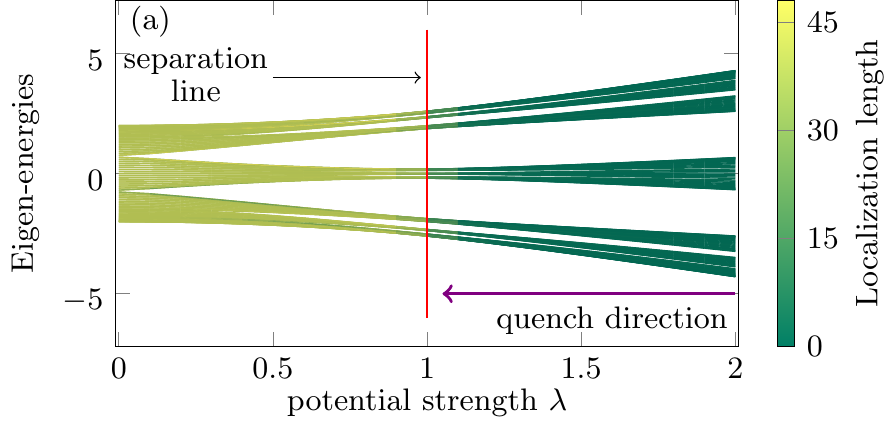}
   \includegraphics[width=\columnwidth,clip=true]{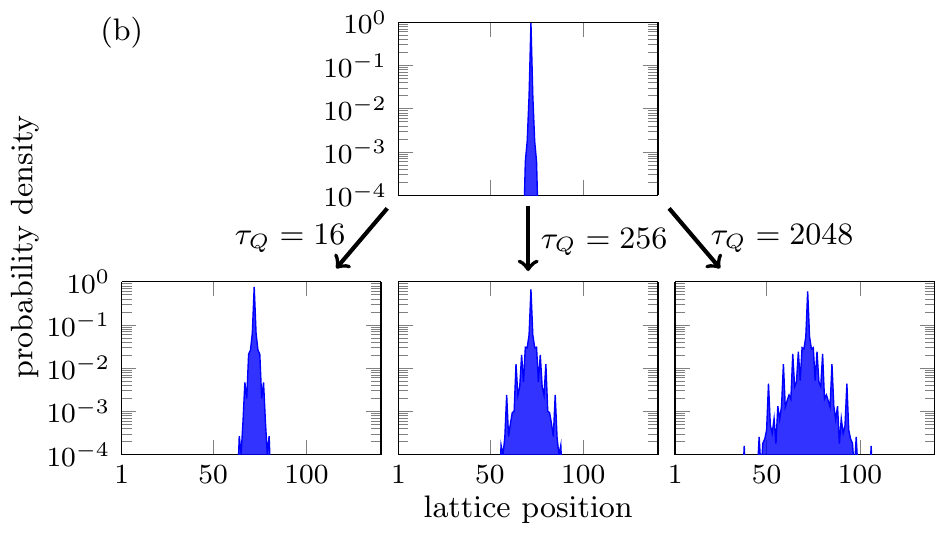}
\end{center}  
\caption{In (a), Eigenstates of AA model are calculated for a lattice of size $L=144$. Color indicates a measure of the localization length of the eigenstate: light green denotes localization length $\approx L/\sqrt{12}$ (de-localized) and dark green denotes localization length of $\mathcal{O}(1)$ (localized). The abrupt color change at critical $\lambda_{c}=1$ separates the localized and de-localized phases. We initialize the quench in the ground state deep in the localized phase and ramp the potential strength $\lambda$ at a finite rate across $\lambda_{c}$. In (b), the top panel shows the initial ground state in the localized phase.
The bottom panels show snapshots of states during a slow ramp through the critical region.
They preserve a finite localization length that grows with a power of the quench time $\tau_Q$.
This scenario could be realized experimentally with a non-interacting Bose-Einstein condensate in a
pseudo-random optical lattice potential as in Ref. \cite{Roati2008}.}
\label{fig:cartoon} 
\end{figure}

A class of 1D models with a quasi-periodic potential can be shown to have a localization-delocalization transition at a critical value of the potential strength. We will focus on one such model: the Aubry-Andr\'{e} (AA) model~\cite{Harper1955, AA1980}. It has attracted a large amount of both theoretical~\cite{Sharma2017, Modugno2009} as well as experimental works~\cite{Roati2008, Lahini2009, quasicrystals}, as it exhibits the key features of localization transition usually manifested by higher dimensional systems, see also Ref.~\cite{MajorMorigi} for a recent proposal. Our main aim here is to investigate a linear quench: starting from the ground state of the system in the localized phase, and subsequently tuning the potential strength adiabatically across the phase transition, see Fig. \ref{fig:cartoon}. Near the critical point, however, due to vanishing relevant energy gap, the dynamics cannot be adiabatic and the system gets excited. This scenario is captured by the quantum version of the Kibble-Zurek mechanism (KZM) which provides a paradigm for describing excitations induced in a system driven through a continuous phase transition at a finite rate. 



 Kibble's scenario was a vision of symmetry breaking thermal transitions in the early Universe \cite{K}. It was substantiated by Zurek's mechanism quantifying adiabaticity of such transitions \cite{Z}. The classical KZM was immediately recognized as a universal theory of the dynamics of phase transitions and verified by numerical simulations \cite{KZnum} and laboratory experiments in various condensed matter experiments \cite{KZexp}. More recently KZM was generalized to quantum phase transitions \cite{QKZ,d2005}. Theoretical works \cite{QKZteor} and experimental tests \cite{QKZexp,deMarco2,Lukin18} followed. A recent experiment with ultracold Rydberg atoms \cite{Lukin18} is an accurate quantum simulation of the exact solution for a linear quench in the quantum Ising chain \cite{d2005}. These days the importance of the quantum KZM stems more from its relevance for adiabatic quantum state preparation rather than cosmology. It quantifies non-adiabaticity of adiabatic evolution from an easy-to-prepare initial ground state to an interesting final one. Its importance as a roadblock for quantum simulation begins to be recognized by  experimental groups \cite{GreinerHubbard}. 

Nevertheless, relatively little attention has been paid to KZM in disordered/localized systems. The experimental work on the Bose glass to superfluid transition \cite{deMarco2} has little overlap with the few theoretical papers on spin chains \cite{random,Roosz2014}. The AA model has a unique potential to bridge this gap. It allows for clear-cut predictions sharing generic features with any localization-delocalization transition. They could be quantum simulated with existing technology. As it is routine nowadays to ramp strength of an optical lattice, a non-interacting Bose-Einstein condensate in a pseudo-random potential \cite{Roati2008} could be easily ramped to the delocalized phase. 

\section{Aubry-Andr\'e model}
The model \cite{AA1980} is defined by a non-interacting Hamiltonian
\bea
H = -\sum_{j=1}^{L}
     c^\dag_{j+1} c_j+{\rm h.c.}
  +  2\lambda\cos\left[2\pi\left(\gamma j+\phi\right)\right] c^\dag_j c_j.
\label{H}
\eea
Here $c^\dag_j$ ($c_j$) are creation (annihilation) operators and $\gamma=(\sqrt5-1)/2$ is an irrational number.
Random $\phi\in[0,1)$ was introduced as a mean to average over the pseudo-random potential.
For periodic boundary conditions, $c_{L+j}=c_j$, the potential must also be periodic,
hence $\gamma$ has to be approximated by a rational number with $L$ in the denominator.
In the following, dependence on $L$ means a sequence $L=F_m$, $\gamma=F_{m-1}/F_m$, 
where $F_m$ are the Fibonacci numbers. 

\begin{figure}[t]
\includegraphics[width=\columnwidth,clip=true]{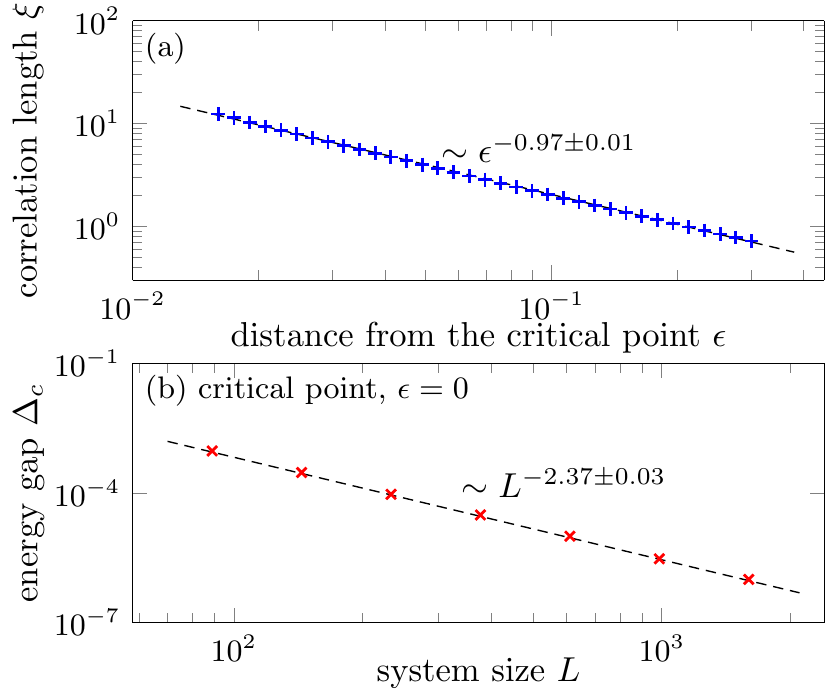}
\vspace{-0.0cm}
\caption{
In (a), 
localization length $\xi$ in function of a distance from the critical point $\epsilon$.
Here $\xi$ was calculated as dispersion of localized ground state on a lattice of $L=987$ sites.
The linear fit $\xi\sim\epsilon^{-\nu}$ yields $\nu=0.97 \pm 0.01$ in good agreement with exact $\nu=1$. 
In (b),
a gap $\Delta_c$ between the ground state and the first excited state at the critical $\lambda=1$ in function of the lattice size $L$. 
Fitting $\Delta_{c}\sim L^{-z}$ yields a dynamical exponent $z=2.37 \pm 0.03$.
All results were averaged over $100$ random~$\phi$. 
}
\label{fig:nu} 
\end{figure}

When $\lambda>1$ all eigenmodes are localized within a localization length $\xi=1/\ln\lambda$. 
Close to the critical point the length diverges as \cite{AA1980},
\be 
\xi\approx\epsilon^{-\nu}, ~ ~ \epsilon = \lambda  - 1,
\label{xi}
\ee
where $\epsilon$ is the distance from the critical point and $\nu=1$ is a correlation-length exponent, see Fig.~\ref{fig:nu}a. 
On the other hand, when $\lambda<1$ the eigenmodes are extended modulated plane waves. 
The model is self-dual \cite{AA1980}. There is a linear map,
\be 
c_{j'}=\frac{1}{\sqrt{L}}\sum_j c_j e^{i2\pi j'(\gamma j+\phi)},
\ee 
that interchanges $\lambda$ and $1/\lambda$ in the Hamiltonian.
It is a symmetry between the localized and delocalized phases.  

The dynamical exponent $z$ can be determined at the critical point from a finite size scaling of energy gap $\Delta_c$ between the ground state and the first excited state:
\be 
\Delta_{c} \sim L^{-z}.
\label{z}
\ee
A linear fit to the log-log plot in Fig.~\ref{fig:nu}b yields $z=2.37\pm0.03$. This value is consistent with the one obtained by 
studies of finite size scaling of superfluid fraction, $z=2.374$\cite{Cestari11}, and very recently by studies of fidelity susceptibility and generalized adiabatic susceptibility, $z=2.375$\cite{Wei19}.

The two exponents, $z$ and $\nu$, are usually enough to determine how the gap opens with the distance from the critical point, namely, $\Delta\sim\epsilon^{z\nu}$. As the gap is zero for $L\to\infty$, this power law may seem not to apply here. However, when replaced by a proper relevant gap, the above relation holds and controls adiabaticity of the evolution.

\section{Kibble-Zurek Mechanism}
By slowly varying $\epsilon$ across the critical point, we want to drive the system (\ref{H}) from an initial ground 
state deep in the localized phase, across the phase transition, into the delocalized phase. 
The question is adiabaticity of this evolution.

\begin{figure}[t]
\includegraphics[width=\columnwidth,clip=true]{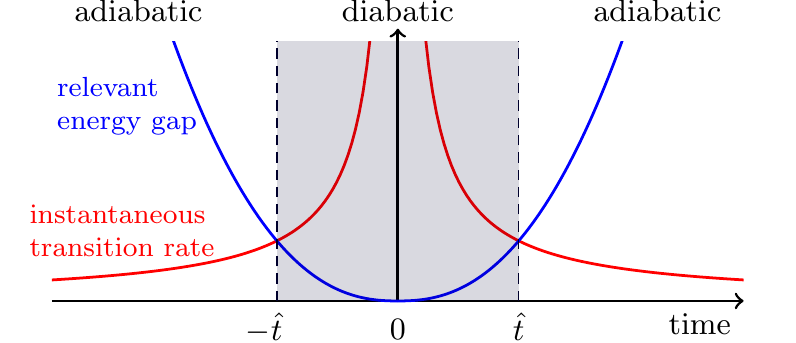}
\vspace{-0.0cm}
\caption{
Linear ramp across the critical point. Here time~$0$ corresponds to the critical Hamiltonian with $\lambda=1$.
The instantaneous transition rate, $\left|\dot{\epsilon}/\epsilon\right| = 1/|t|$, diverges at the critical point
and, at the same time, the relevant energy gap closes like $\sim|\epsilon|^{z\nu}$. Consequently, in the neighborhood
of the critical point, between $-\hat t$ and $\hat t$, the evolution is not adiabatic.
}
\label{fig:KZcartoon} 
\end{figure}

Near the critical point generic $\epsilon(t)$ can be linearized:
\bea 
\epsilon(t) \approx -t/\tau_Q.
\eea
Its slope is determined by a quench time $\tau_Q$. When the gap relevant for excitations opens with the distance to the critical point like $\Delta\sim|\epsilon|^{z\nu}$, then the evolution must be adiabatic sufficiently far from the critical point.
It crosses over to diabatic evolution when the instantaneous transition rate,
$ 
\left|\dot{\epsilon}/\epsilon\right| = 1/|t|,
$
equals the relevant energy gap
$
\Delta(t) \sim \left|t/\tau_Q\right|^{z\nu}.
$
The two are equal at crossover times $t = \pm \hat t$, where
\be
\hat t \sim {\tau_Q}^{z\nu/ (1+z\nu)},
\label{hatt}
\ee
see Fig.~\ref{fig:KZcartoon}.
At $-\hat t$ the state is still the adiabatic ground state at $-\hat\epsilon$, where
$
\hat \epsilon=\hat t/\tau_Q\sim{\tau_{Q}}^{-1/(1+z\nu)},
$  
with a localization length 
\bea 
\hat \xi \sim \hat\epsilon^{-\nu} \sim \tau_{Q}^{\nu/(1+z\nu)}.
\label{hatxi}
\eea
This KZ length is a characteristic scale of length just as $\hat t$ is a characteristic scale of time.

In zero order impulse approximation, this state \textit{freezes out} at $-\hat t$ and does not change until $\hat t$ when the evolution becomes adiabatic again. At $\hat t$ the frozen state is no longer the ground state but an excited state with a localization length $\hat\xi$. It is an initial state for the adiabatic process that follows after $\hat t$. 
As a result of non-adiabaticity, the wave-packet does not follow the adiabatic ground state -- whose localization length would diverge at the critical point -- but enters the delocalized phase with a finite localization length $\hat\xi$. When the quench time $\tau_Q\to\infty$, then the localization length diverges to infinity and the adiabatic limit is recovered.  

\begin{figure}[t]
\includegraphics[width=\columnwidth,clip=true]{{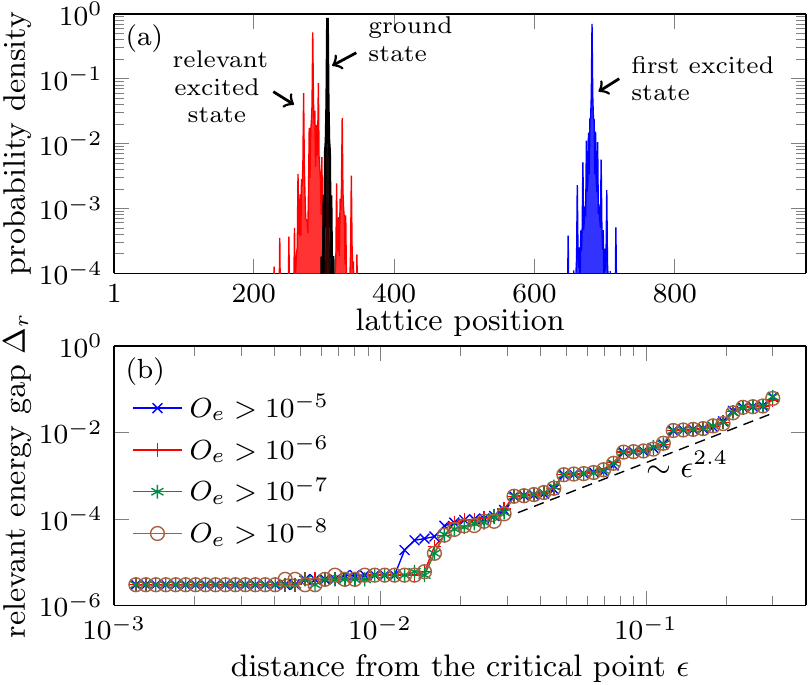}}
\vspace{-0.0cm}
\caption{
In (a) we show localized eigenstates at $-\hat t$: 
the adiabatic ground state (black),
the relevant excited state with the most probable transition (red),
and the first excited state (blue).
The relevant state overlaps with the ground state.
The first excited state, even though it has the smallest gap,
has no overlap with the ground state and hence is irrelevant for excitation.
Consequently, in (b) we plot the relevant gap $\Delta_r$ as a function of the distance from the critical point $\epsilon$. The four plots correspond to different choices of pre-selecting the relevant states with $O_e$ greater than $10^{-5},10^{-6},10^{-7},10^{-8}$. These log-log plots have a common step-like structure with an overall linear dependence on $\epsilon$. Fit to the linear region yields $\Delta_r\sim\epsilon^{z\nu}$ with $z\nu=2.4 \pm 0.1$ close to the expected $z\nu\simeq2.37$ obtained from the finite size scaling at criticality, see Fig.~\ref{fig:nu}b. We focus on the linear region as finite size effects, shown in Fig.~\ref{fig:nu}b, appear below $\Delta \sim 10^{-5}$.  Here $\Delta_r$ was averaged over $100$ random $\phi$ on a lattice of $L=987$ sites.
} 
\label{fig:relevant} 
\end{figure}

\section{Relevant Gap}
The localized phase is gapless but eigenstates are localized within the finite localization length $\xi$. 
In the adiabatic perturbation theory, a necessary condition to transfer from the ground state $|g\rangle$ 
to an excited state $|e\rangle$ is that the two states overlap. The gap relevant for adiabaticity 
is the gap between the ground state and the first excited state whose support overlaps with the ground state, see Fig.~\ref{fig:relevant}(a). 
In analogy to the excited states confined within a finite lattice, see Eq.~(\ref{z}), the excited states that are 
confined within the localization length $\xi$ of the ground state should be separated from the ground state
by a relevant gap
\be 
\Delta_r \sim \xi^{-z} \sim \epsilon^{z\nu} = \epsilon^{z}.
\label{Deltar}
\ee
This prediction is consistent with numerics in Fig.~\ref{fig:relevant}b.

A more precise definition of the relevant gap and relevant states follows from the adiabatic perturbation theory, 
where the transition rate from $|g\rangle$ to $|e\rangle$, separated by a gap $\Delta_e$, depends on 
a matrix element \cite{Messiah}
\be 
O_e=|\langle e|dH/d\epsilon|g\rangle|.
\ee 
The relevant states are those with non-zero $O_e$.
In practice we use a cut-off, say $|O_e|>10^{-8}$, but the general conclusion does not depend on its exact value.
Its gap is the relevant gap $\Delta_r$ shown in Fig.~\ref{fig:relevant}b. 
A state with the most probable transition is the one with maximal $O_e/\Delta_e^2$. 
\begin{figure}[t]
\includegraphics[width=\columnwidth,clip=true]{{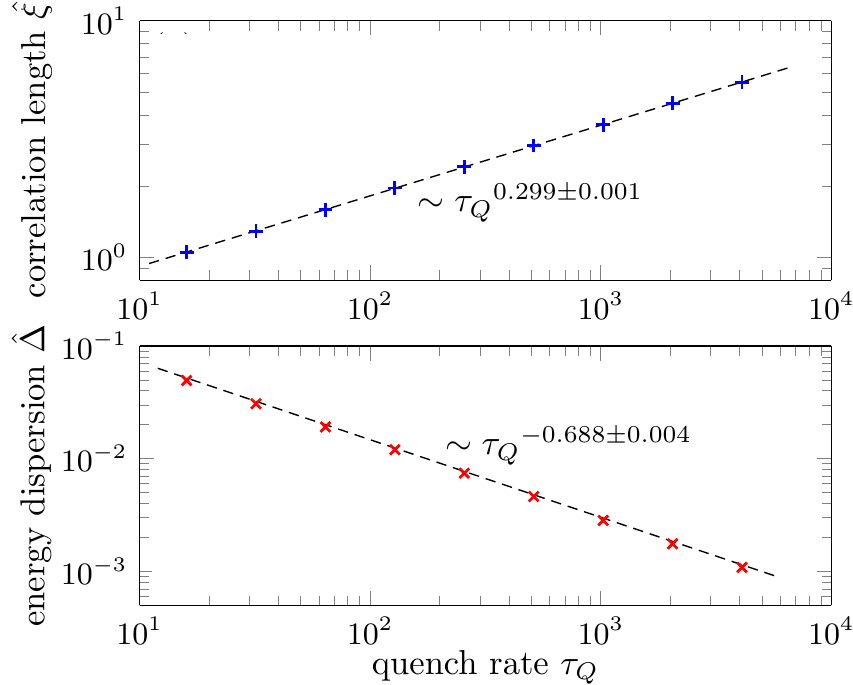}}
\caption{
In (a),
the width of the wave-packet at the critical point as a function of the quench time $\tau_Q$.
Fit gives $\hat\xi\sim \tau_Q^{0.299\pm0.001}$, compare with Eq.~(\ref{hatxi}). 
In (b), similar plot for the energy dispersion at the critical point. The linear fit yields $\hat\Delta\sim\tau_Q^{-0.688\pm0.004}$, see Eq.~(\ref{hatDelta}).
Here the lattice size $L=987$ and averaging is done over $10$ random values of $\phi$.}
\label{fig:quench} 
\end{figure}

\section{KZ Power Laws}
We tested our predictions with numerical simulations. We used a smooth tanh-profile
$
\epsilon(t) = - \tanh\left(t/\tau_Q\right)
$
starting from $t_i=-5\tau_Q$ in order to suppress excitation originating from the initial discontinuity of
the time derivative $\dot\epsilon$ at $t_i$. 

The evolution becomes diabatic at $-\hat t$ when the localization length is $\hat\xi$, see Eq.~(\ref{hatxi}). 
This $\hat\xi$ is expected to characterize the width of the wave packet in the diabatic regime between $-\hat t$ 
and $\hat t$. Indeed, in Fig.~\ref{fig:quench}a we plot the width at the critical point -- estimated as dispersion 
of the probability distribution -- in function of $\tau_Q$. The power-law fit and Eq.~(\ref{hatxi}) imply $z=2.34 \pm 0.01$ for the exact $\nu=1$ and $z=2.31\pm0.02$ for $\nu\simeq0.97$ estimated in Fig.~\ref{fig:nu}.

The relevant states that become excited during the evolution are separated from the adiabatic ground state by 
the relevant gap in Eq.~(\ref{Deltar}). At $-\hat t$ the relevant gap is
\be 
\hat\Delta \sim \hat\xi^{-z} \sim \tau_Q^{-z\nu/(1+z\nu)}.
\label{hatDelta}
\ee
This energy scale is expected to characterize energy dispersion of the excited state in the diabatic regime
between $-\hat t$ and $\hat t$. We test this prediction in Fig.~\ref{fig:quench}b where we plot the energy dispersion
at the critical point as a function of $\tau_Q$. The power-law fit and Eq.~(\ref{hatDelta}) imply $z=2.21 \pm 0.04$ for the exact $\nu=1$ and $z=2.27\pm0.05$ for the fitted $\nu\simeq0.97$.

The values of $z=2.31$ and $z=2.27$ obtained, respectively, from $\hat\xi$ and $\hat\Delta$ setting $\nu=0.97$ 
differ by $1\%$. Their average is $3\%$ below $z\simeq2.37$ yielded by the finite size scaling in Fig.~\ref{fig:nu}.
Given that similarly estimated $\nu\simeq0.97$ is $3\%$ below the exact $\nu=1$,
the discrepancies are comparable to this systematic error. 
Therefore, within small error bars, our numerical results are consistent with the predictions.  

\begin{figure}[!t]
\includegraphics[width=\columnwidth,clip=true]{{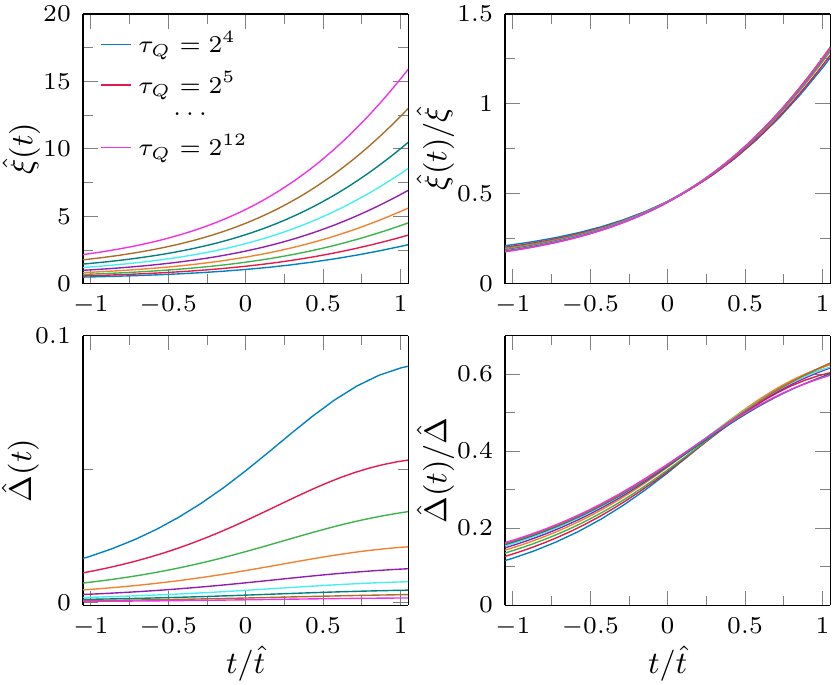}}
\caption{
The top/bottom row shows the width/dispersion in function of scaled time.
In the left panels the two quantities are not scaled.
In the right panels they are scaled and collapse to their respective scaling functions.
The collapse demonstrates the KZ scaling hypothesis.
}
\label{fig:hyp}
\end{figure}

\section{KZ scaling hypothesis}
In the diabatic regime, between $-\hat t$ and $\hat t$, $\hat\xi$ and $\hat\Delta$ are the relevant scales of 
length and energy, respectively. They diverge in the adiabatic limit, $\tau_Q\to\infty$, where they become
the only relevant scales in the long wavelength and low frequency regime. This logic justifies 
the KZ scaling hypothesis \cite{KZscaling} for a correlation length $\hat\xi(t)$ and energy dispersion 
$\hat\Delta(t)$ in the diabatic regime:
\bea 
\hat\xi(t)    = \hat\xi    F_{\xi}\left(t/\hat t\right),~~~
\hat\Delta(t) = \hat\Delta F_{\Delta}\left(t/\hat t\right),
\eea 
where $F_\xi$ and $F_\Delta$ are two non-universal scaling functions. 
It is confirmed by the collapse of 
scaled plots in Fig.~\ref{fig:hyp}. This naturally includes scaling of the width of wave packets at $t=+\hat t$ shown in the lower panels of Fig.~\ref{fig:cartoon}. 

\section{Discussion}This work demonstrates that a linear ramp across the localization-delocalization transition is not adiabatic
and results in a final excited state with a finite localization length and non-zero energy dispersion. 
These two quantities satisfy power laws with respect to the ramp time $\tau_Q$ with the universal Kibble-Zurek exponents.

The same scenario is expected to apply more generally to a ramp across a mobility edge in higher dimensional disordered systems. To test this hypothesis, we choose the generalized Aubry-Andr\'{e} model\cite{Sarma15} (GAA) which possess an energy-dependent mobility edge similar to the 3D disordered systems. Appendix \ref{appa} is devoted to the scenario when we quench across the mobility edge in GAA . Results akin to the present work has been obtained. This is expected since the ordinary AA model is just a special case of the generalized AA model where the mobility edge coincides with the critical point. Thus our relevant gap hypothesis has been established on a more general and stronger footing.

We expect that KZM also applies to many-body localization-delocalization transitions like
the Bose glass to superfluid quantum phase transition in the random Bose-Hubbard model \cite{deMarco2} or many-body localization-delocalization of all eigenstates \cite{Basko2006}.
In the latter, in the Heisenberg picture
we expect the localized integrals of motion to freeze out at the length $\hat\xi$ when ramped across the critical point. The implication of the KZM in driven AA systems\cite{Flach2014}, where it has been proved that the driving terms can alter the localization properties in the insulating phase and induce delocalization, can be studied with the help of numerical approaches similar to those used in this paper.

Therefore, our example is a representative toy model of the Kibble-Zurek mechanism in a localization-delocalization transition.
%

\section*{acknowledgements}
This research was funded by National Science Centre (NCN), Poland under projects 2016/23/B/ST3/00830 (JD) and 2016/23/D/ST3/00384 (MMR), and NCN together with European Union through QuantERA ERA NET program 2017/25/Z/ST2/03028 (AS).


\appendix
\section{Generalized Aubry-Andr\'e Model}
\label{appa}

The so-called generalized Aubry-Andr\'e (GAA) model is obtained by replacing the potential term of the AA Hamiltonian Eq.~(\ref{H}) by $ 2\lambda\frac{\cos\left[2\pi\left(\gamma j+\phi\right)\right]}{1-\alpha\cos\left[2\pi\left(\gamma j+\phi\right)\right]} c^\dag_j c_j $. It possesses an energy-dependent mobility edge separating the localized and delocalized phases. Reference\cite{Sarma15} provides an analytical expression relating the mobility edge, $E$, with the potential strength $\lambda$:
\be 
\alpha E = 2 (1 - |\lambda|) {\rm sign}(\lambda).
\ee 
This brings the model closer to the more generic 3D Anderson model which also has an energy-dependent mobility edge. As a self-consistency check, notice that for $\alpha = 0$ the mobility edge reduces to $\lambda = 1$ which is precisely the energy-independent critical point of the AA model. 
\begin{figure}[b]
\includegraphics[width=\columnwidth]{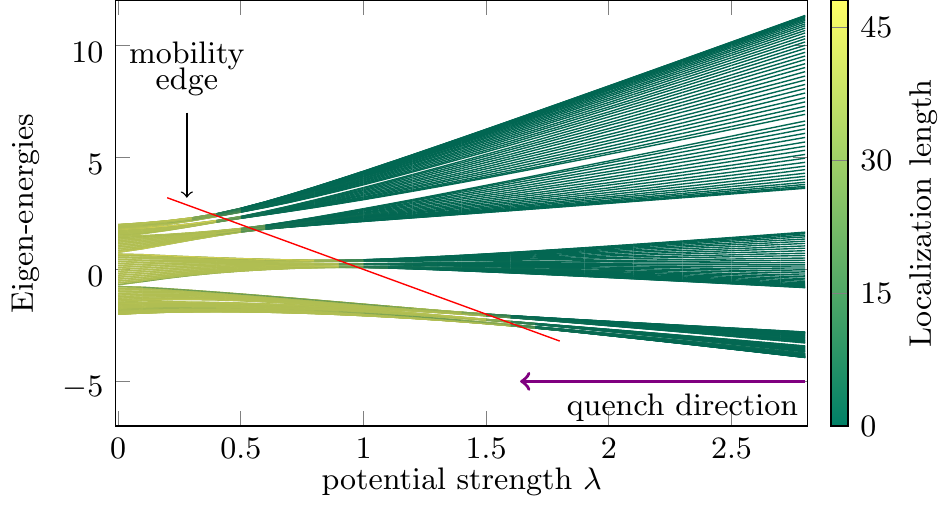}
\caption{
Eigenstates of GAA model are calculated taking $\alpha=0.5$ for a lattice size $L=144$. Color indicates a measure of the localization length of the eigenstate: light green denotes localization length $\approx L/\sqrt{12}$ (completely de-localized) and dark green denotes localization length of $\mathcal{O}(1)$ (completely localized). The abrupt color change separates the localized and de-localized phases defining the mobility edge (red line), which is consistent with the analytic prediction.
We initalise the quench in the ground state deep in the localized phase and ramp the potential strength $\lambda$ at a finite rate across the 
mobility edge.}
\label{fig:mobedge}
\end{figure}

\begin{figure}[t]
\includegraphics[width=\columnwidth,clip=true]{{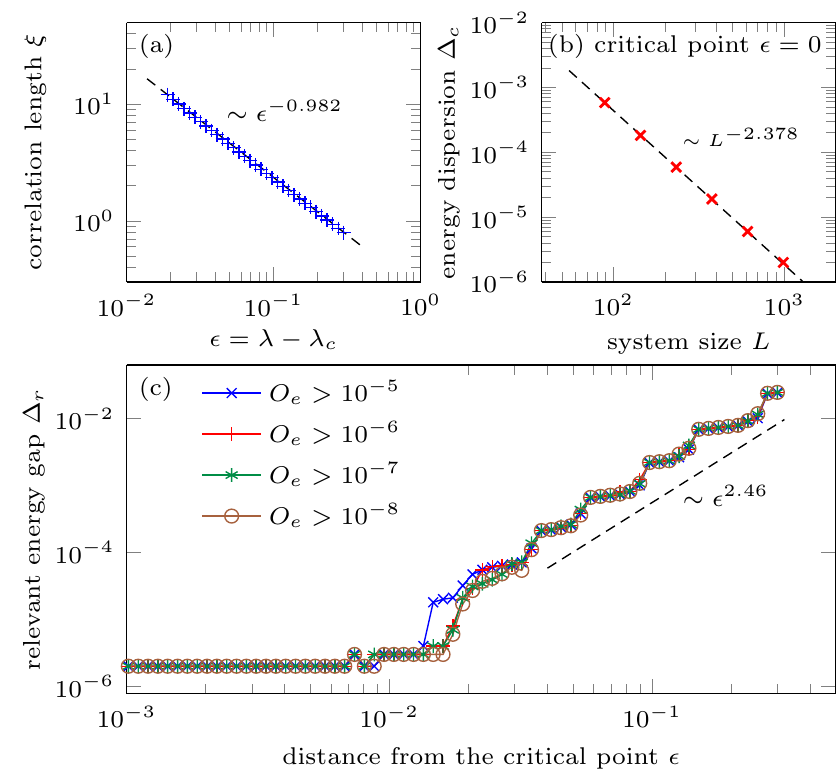}}
\caption{
Correlation and dynamical exponents for the generalized Aubry-Andr\'{e} model:
In (a),
the localization length $\xi$ in function of a distance from the critical point $\epsilon$. Here $\xi$ was calculated as dispersion of a localized ground state on a lattice of $L=987$ sites. The fit $\xi\sim\epsilon^{-\nu}$ yields $\nu=0.982$ similar to $\nu=0.97$ obtained for the AA model, see  Fig.~\ref{fig:nu}.
In (b),
a gap $\Delta_c$ between the ground state and the first excited state at the critical point $\lambda_{c} \approx 1.638$ in function of the lattice size $L$. 
Fitting $\Delta_{c}\sim L^{-z}$ yields a dynamical exponent $z=2.378$ which is, again, similar
to the AA model(see  Fig.~\ref{fig:nu}).
Results were averaged over $100$ random~$\phi$. 
In (c), 
we plot the relevant gap $\Delta_r$ as a function of the distance from the critical point $\epsilon$ 
similar to Fig.~\ref{fig:relevant} for the AA model. Once again the fit to the linear region yields $\Delta_r\sim\epsilon^{z\nu}$ with $z\nu=2.46$ close to the expected $z\nu\simeq2.378$ obtained from the finite size scaling at criticality, see Fig.~\ref{fig:nu}b. We focus on the linear region as finite size effects,  shown in Fig.~\ref{fig:nu}b, appear below $\Delta \sim 10^{-5}$. Here $\Delta_r$ was averaged over $100$ random $\phi$ on a lattice of $L=987$ sites.}
\label{fig:staticgaah}
\end{figure}
\begin{figure}[b]
\includegraphics[width=\columnwidth,clip=true]{{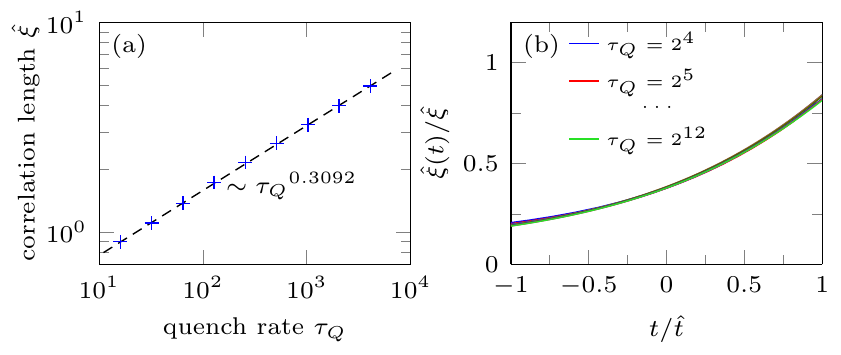}}
\caption{
In (a), 
the width of the wave-packet at the critical point $\lambda_c$ as a function of the quench time $\tau_Q$. The fit gives $\hat{\xi} \sim {\tau_Q}^{0.3092}$ which yields $z=2.27$ for $\nu = 0.982$.
In (b), 
the scaled widths in function of the scaled times for different $\tau_Q$'s collapse to a common scaling function. The collapse demonstrates the KZ scaling hypothesis.
}
\label{fig:quenchgaah}
\end{figure}
We proceed to investigate the adiabatic quench by slowly tuning the potential strength $\lambda>0$ across the mobility edge $\lambda_c$. 
For definiteness, we choose a generic $\alpha = 0.5$, which results in a mobility edge $E = 4(1 - \lambda)$. We verify this numerically by calculating the localization length of the eigenstates of the GAA model for varying value of the parameter $\lambda$, see Fig.~\ref{fig:mobedge}.
Here, the correlation length is used as a parameter to determine whether a specific eigenstate is localized or not.

Now we verify the static scaling results by following the same methods as used in the AA model. 
We focus on the ground state whose mobility edge is at
\be 
\lambda_c\approx 1.638
\ee 
and measure the distance from the critical point as $\epsilon=\lambda-\lambda_c$. 
Fig.~\ref{fig:staticgaah} shows that the obtained correlation and dynamical exponents 
are close to the corresponding ones for the AA model. \\
Finally, we test those static predictions against the results obtained from numerical simulation of a quench, where akin to the main text we
use $tanh$ profile to suppress excitations appearing at the begining of the evolution.
The evolution is initialized in the ground state at $\epsilon=0$.
In Fig.~\ref{fig:quenchgaah}a we show the width of the evolving wave-packet at the critical point - estimated as dispersion of the probability distribution - in function of $\tau_Q$. The power law fit implies $z = 2.27$ for $\nu = 0.982$. We also verify the KZ scaling hypothesis in Fig.~\ref{fig:quenchgaah}b.

\end{document}